# Research on CPI Prediction Based on Natural Language Processing

Xiaobin Tang [1] Nuo Lei[2]*

[1] [2]  University of International Business and Economics

*    Corresponding author

E-mail address:  tangxb@uibe.edu.cn (Xiaobin Tang)

201857011@uibe.edu.cn (Nuo Lei)

**Funding:** This work was supported by CAAI-Huawei MindSpore Open Fund [Grant Numbers: No. CAAIXSJLJJ-2021-045A] and National Social Science Foundation of China [No. 22&ZD164]

**Abstract:** In the past, the seed keywords for CPI prediction was often selected based on empirical summaries of research and literature studies, which were prone to select omitted and invalid variables. In this paper, we design a keyword expansion technique for CPI prediction based on the cutting-edge NLP model, PANGU. We improve the CPI prediction ability using the corresponding web search index. Compared with the unsupervised pre-training and supervised downstream fine-tuning natural language processing models such as BERT and NEZHA, the PANGU model can be expanded to obtain more reliable CPI-generated keywords by its excellent zero-sample learning capability without the limitation of the downstream fine-tuning data set. Finally, this paper empirically tests the keyword prediction ability obtained by this keyword expansion method with historical CPI data.

# 1. Introduction

Macroeconomic indicators reflect the main trend of the national economy and have important reference significance for the formulation of government fiscal and monetary policies. Among them, the consumer price index (CPI) is one of the important macroeconomic indicators. On the one hand, it reflects the living cost of ordinary people and is closely related to the daily life of the people. On the other hand, it is also an important reference for the Chinese growing internal economic cycle. Therefore, using statistical methods to predict CPI has become a research focus in academia and industry under the background of expanding domestic demand and establishing dual circulation economic development.

At present, the statistics of macroeconomic indicators represented by CPI mainly come from on-the-spot visits of investigators. The obtained macroeconomic indicator data inevitably has a time lag problem, which affects the prediction effect of the model. In the era of big data, the rise of big data, the Internet, artificial intelligence, and other technologies has provided massive research data for CPI forecasting. Among them, the network search index is composed of the search frequency of people in search engines for some specific keywords, which can reflect people's attention to specific items in real time and has strong timeliness. Therefore, the introduction of keywords into the network search index can greatly improve the real-time prediction effect of traditional methods for CPI indicators, and has been widely used in research. However, the selection of specific keywords is limited by the experience and subjective judgment of researchers, and there are often omissions to a considerable extent, which will affect the prediction effect of the final model on CPI. In view of the above problems, this paper introduces natural language processing (NLP) to expand keywords. Based on subjectively selected seed words, NLP technology is used to expand more keywords with good predictive ability and their corresponding network search index, thereby improving the forecasting effect of CPI. Therefore, this paper considers using natural language processing technology to obtain more keywords with excellent predictive ability based on some seed keywords.

The structure of this paper is as follows: Chapter 2 reviews the literature on using search index to improve CPI forecasting effect; Chapter 3 focuses on the NLP model represented by PANGU; Chapter 4 explains the expansion and application of PANGU model to specific keywords and CPI forecasting; Chapter 5 empirically tests the keyword expansion model proposed in this paper through historical CPI and keyword search index. The sixth chapter summarizes the research of this paper.

# 2. Literature Review

The search indexes of keywords are widely used in CPI related prediction. Liu et al. (2014) uses Baidu search index of auto recommended keywords to study the connection between Xinjiang CPI and public attention represented by Baidu index, resulting a good prediction with only 4.65% predicted average absolute error. Skipper and Andrea (2015) use Google Trends of *ex ante* keywords to do nowcasting price prediction in Central America, and find out that the introduction of Internet search indexes improve the prediction performance. Xin et al. (2015) construct keywords base from CPI related news and choose the Google search data of nouns and their synonyms to do the prediction by MIDAS model. Liu and Zhang (2019) select keywords from eight CPI categories and subjective keywords popular on the Internet to predict CPI half a month before official release. It can be found that although various scholars have introduced the CPI-related keyword search index in their research to improve the prediction effect, their selection criteria are different. It not only relies mostly on the subjective judgment of the researchers, but also leads to omissions and errors of keywords.

With the development and progress of artificial intelligence technology, the ability of natural language processing models to understand texts has been continuously improved. At present, the Transformer structure proposed by Vaswani et al. (2017) [6] has been used in multiple NLP models, and even exceeded human performance in various test tasks. Thus, it has been widely used in the field of economic research. For macroeconomic indicators such as CPI prediction keyword expansion task, the NLP model can quickly process massive texts, so as to efficiently expand on manually selected seed keywords, and improve the model's ability to predict CPI. Currently, it is gradually attracting research attention. Xiaobin Tang(2021) [7] used the human-machine interactive TF-IDF algorithm and the BERT model to expand the 84 keywords related to CPI to obtain more keywords with predictive ability, and empirically tested the prediction ability of keywords obtained by the expansion. Among them, the BERT model performs downstream fine-tuning on the LCQMC dataset with massive pre-training, so as to judge whether the semantics of the two words are the same. However, the model effect obtained by downstream fine-tuning depends to a large extent on the selection of datasets, and different datasets may lead to significant differences in the results of expanded keywords. On the other hand, there are few Chinese keyword datasets, and most of them are aimed at specific fields such as news and finance. Therefore, it is difficult to obtain suitable keyword datasets for downstream fine-tuning to solve specific CPI keyword expansion problems.

Large-scale pre-trained language models (PLMs) represented by the PANGU model provide a solution to the downstream fine-tuning dataset selection problem such as the BERT (Zeng et al., 2021) [8]. Through super-large-scale pre-training and model parameters, PLMs have superior model performance in few-shot learning and even zero-shot learning problems, which can get rid of the constraints of downstream fine-tuning datasets. As an excellent Chinese PLMs model, the PANGU model has obtained lots of parameters through training on high-quality Chinese text data, so that it can achieve outstanding performance in a series of small-sample learning and zero-sample learning problems. In the prediction keyword expansion problem of CPI, the superior zero-shot learning ability of the PANGU model can judge and expand keywords in multiple domains without a given fine-tuning dataset. Therefore, this paper uses the PANGU model to expand the CPI prediction keywords, which makes the expansion process of the keywords more reasonable.

The specific research route of this paper is as follows. Firstly, starting from the internal composition and external influencing factors of the CPI index, appropriate keywords are selected through literature summarization from both microscopic and macroscopic perspectives to form the initial seed keywords. Second, we use the web crawler to obtain Baidu Baike text related to CPI. Third, natural language processing models such as BERT model and PANGU model are used to extract keywords based on massive encyclopedia texts, so as to expand potential predictor variables. Then we use web crawlers to obtain the Baidu Index of keywords from search engines. Finally, this paper uses PCA and regression analysis to empirically test the prediction effect of the expanded keyword search index, so as to compare and judge the prediction effect of keywords obtained through NLP expansion

## 3. Models and Algorithms

This paper mainly uses the PANGU model to expand new high-quality keywords based on the summarized CPI seed keywords. The PANGU model is a large-scale pre-training language model for Chinese developed by Huawei Peng Cheng Lab, first developed under MindSpore framework[9], which is benchmarked against GPT-3 (Brown et al., 2020)[10]. It has good performance in natural language generation tasks (NLG) and natural language understanding tasks (NLU). Especially in multiple few-shot and zero-shot tasks, it even surpasses the natural language model after downstream tuning of the corresponding dataset, such as BERT, NEZHA, etc. The GPT-3 developed by OpenAI based on the Transformer structure is mainly based on English text for pre-training, so it cannot handle NLP tasks for Chinese well. Compared with the GPT-3 model, the PANGU model is developed for Chinese natural language processing tasks and pre-trained on a large number of Chinese texts, so it can better understand the meaning of Chinese texts than the GPT-3 model. This could help the model to better extract high-quality keywords related to CPI. On the other hand, the PANGU model code has been open-sourced, so that it can be employed more conveniently.

Compared with the Transformer Encoder structure used in BERT (Devlin et al, 2018)[11], the main structure of the PANGU model is based on a multi-layer Transformer decoder (Decoder) and a Query layer structure superimposed on it. Therefore, BERT is more suitable for dealing with vectorized representation problems, while the PANGU model is more suitable for dealing with text generation models. Here we would introduce the structure of PANGU based on the research of Zeng et al. (2021)[8] and Devlin et al. (2018)[11]. For the CPI prediction keyword expansion problem dealt with in this paper, the PANGU model will perform the following encoding preprocessing on the input CPI-related text, so as to convert it into a numerical form that the Transformer can process. For a piece of input text $T = t_1, t_2, \ldots, t_n$ of length n, the model first converts the input text into a matrix $X_0$ through Token Embedding and Position Embedding, so as to perform subsequent calculations. The calculation formulas of Token Embedding and Position Embedding are as follows:

$$X_0 = T W_{Text} + W_{Pos} \tag{1}$$

The obtained matrix $X_0$ is the initial encoding result of the CPI-related text, which contains all the information of the input text. In order to obtain CPI-related keywords with excellent predictive ability, the model needs to be able to fully understand CPI-related texts through a multi-layer network. Based on the matrix, the PANGU model uses a multi-layer Transformer Decoder to calculate $X_l$, and the calculation formula is as follows:

$$X_l = TransformerDecoder(X_{l-1}) \tag{2}$$

Among them $l \in [1, L]$, it represents that the PANGU model could deeply understand the CPI-related text through the Transformer decoder structure of $L$ layers, which helps to extract effective CPI prediction keywords.

Transformer is the most important structure in the PANGU model, mainly composed of two sub-layers. The first sub-layer is a multi-head self-attention mechanism (MHA), which can understand the key sentences in the text in a targeted manner by stacking multiple self-attention mechanisms. The second sublayer is a fully connected Feed-Forward Network (FFN). Before the multi-head self-attention mechanism and the feed-forward neural network, the pre-layer normalization mechanism was adopted to accelerate the training speed of the Transformer and help the model to mine the text of keywords later.

Specifically, the multi-head self-attention mechanism is an important basis for the PANGU model to fully understand the deep meaning of text. Through the weighted calculation in the self-attention mechanism, the PANGU model can effectively solve the long-distance dependency problem that other natural language models cannot handle. In the CPI keyword expansion problem, the texts containing high-quality predicted keywords are often very long. Therefore, traditional natural language processing models often cannot effectively understand the meaning of the text, so it is naturally difficult to extract suitable keywords. The self-attention mechanism can well grasp the global information of the text through long-distance dependencies, and focus on the important position that needs more attention. Therefore, the PANGU model is very suitable for keyword expansion for the CPI prediction problem. At the same time, the multi-head model design in the PANGU model can achieve effective parallel stacking, and then extract effective information from multiple semantic dimensions in the text. For the keyword expansion problem of CPI prediction problem, the mechanism can deeply understand the semantics of keywords in the original text, and thus expand to obtain keywords with excellent prediction ability for CPI. The specific calculation method is as follows. First, the original encoding result is converted through the three weight matrices of query $Q$, key $K$ and value $V$. The formula is:

$$Q = W^q X_{l-1} \tag{3}$$

$$K = W^k X_{l-1} \tag{4}$$

$$V = W^v X_{l-1} \tag{5}$$

The attention score combines all the information of matrix Q, K, and V with the scale factor $\sqrt{d}$ and softmax function. The scaling processing is to prevent the gradient explosion problem caused by too large values. The calculation formula is as follows:

$$Attention(X_{l-1}) = Softmax\left(\frac{Q \times K'}{\sqrt{d}}\right)V \tag{6}$$

The multi-head attention mechanism adds up the attention scores of each dimension by stacking multiple attention heads, and obtains text information in multiple semantic spaces from the summary. Therefore, for the multi-head attention mechanism of N heads, the calculation formula is as follows:

$$MultiHeadAttention(X_{l-1}) = \sum_N Attention(X_{l-1})W^m \tag{7}$$

Here $W^m$ is also the weight matrices to be trained. On this basis, the PANGU model adds a pre-layer normalization mechanism, so the final output of the multi-head attention mechanism is as follows:

$$X_l^{MHA} = X_{l-1} + MultiHeadAttention(LayerNorm(X_{l-1})) \tag{8}$$

The PANGU model further mines the information contained in the multi-head attention mechanism through the FFN layer. The GeLU function is used as the activation function of the model. The formula is as follows:

$$X_l = X_l^{MHA} + FFN\left(LayerNorm(X_l^{MHA})\right) \tag{9}$$

Among them, FFN is a fully connected neural network, and LayerNorm is the same as the self-attention mechanism, which is normalized before the layer.

On top of the above Transformer structure, the PANGU model also designs a Query layer to explicitly generate the expected output results. The structural design of the Query layer is basically the same as that of the Transformer layer, but its query matrix has an additional layer of encoding to mark the position of the next predicted word. For the output of the last layer of the Transformer structure $X_L$, the query matrix Q is calculated differently as follows:

$$Q = W_n^{q \times p} \tag{10}$$

Based on the above structure, the PANGU model has an excellent performance on a variety of cross-domain Chinese natural language processing tasks on the basis of massive pre-training, and is suitable for the keyword expansion task of CPI.

## 4. Experimental Design

Based on the seed keywords summarized by experience, this paper crawls the CPI-related web texts in Baidu Baike through web crawler and constructs a potential CPI prediction keyword database. In the next step, this paper uses the zero-shot capability of the PANGU model pre-trained in massive texts to process CPI-related Baidu Baike texts to obtain high-quality CPI prediction keywords. Finally, this paper uses the methods of Pearson correlation coefficient and principal component analysis to screen the expanded CPI prediction keywords, so as to further improve the final prediction ability of the CPI series.

**4.1. PANGU Pre-training**

In the past, BERT and NEZHA models based on the Transformer structure divided the natural language processing task into two stages: unsupervised pretraining and supervised downstream fine-tuning. This model design requires targeted downstream fine-tuning for each specific downstream task. For the CPI prediction keyword expansion problem, the corresponding datasets collected for CPI are quite scarce, which greatly limits the usage of the models. With the continuous improvement of computing ability, researchers found that the increase of model parameters can significantly improve the performance of the model in various downstream tasks. So with a large amount of unsupervised pre-training, the model can have great performance in various downstream tasks. The performance is close to or even better than previously fine-tuned models. Therefore, the realization of general artificial intelligence through ultra-large-scale pre-training and model parameters has become the frontier direction of natural language processing. The PANGU model is representative of the super large model in Chinese natural language processing. Its 200 billion parameters and 1.1TB high-quality Chinese corpus make it have good performance in various Chinese NLP tasks.

In terms of unsupervised pre-training datasets, the PANGU model collects a large amount of Chinese corpus to help the model better understand Chinese text processing tasks and improve the generalization ability of the model. The specific data sources include a large number of Chinese open source data sets, crawler data of Chinese web pages, encyclopedia texts and e-books, etc., and have undergone a series of data cleaning and deduplication processing to ensure the high quality and unbiasedness of the data set. This in turn avoids negative impacts on downstream tasks. At the same time, since the encyclopedia text is learned and processed in the pre-training, the model is more suitable for the model design of the Baidu Baike text corresponding to the CPI keywords in this paper, so that the Zero-shot CPI keyword augmentation can be achieved without fine-tuning of the downstream data set.

In terms of the specific pre-training tasks of the model, the PANGU model refers to the design idea of the GPT series model to autoregressively predict the next word of the text sentence in the dataset (Radford et al, 2018)[12], so as to achieve unsupervised model training. Assuming that the text sequence $T = \{t_1, t_2, ..., t_N\}$ contains $N$ characters (tokens), the PANGU model will autoregressively infer the correct character answer for the next position based on all previous characters, thereby maximizing the log-likelihood function as follows:

$$L = \sum_{n=1}^{N} log\, p(t_n|t_1, ..., t_{n-1}; \theta) \tag{11}$$

$\theta$ represents the internal parameters of the PANGU model, which will be estimated by back-propagation to minimize the likelihood function in specific training. After obtaining the corresponding model parameters through many training estimates, the PANGU model can perform inference for downstream tasks in multiple domains including CPI.

### 4.2. PANGU Zero-shot Learning

The biggest advantage of PANGU model pre-training with a large number of samples is its outstanding performance in small-shot learning and zero-shot learning problems. Due to the rich pre-training parameters, the meaning of the text has been fully understood. This enables the PANGU model to avoid the model bias caused by the selection of inappropriate downstream fine-tuning datasets through zero-sample learning in the expansion of CPI prediction keywords, thereby increasing the credibility of the extracted keywords.

In this paper's CPI keyword expansion problem, it is necessary to use the PANGU model to extract appropriate keywords from Baidu Baike related to the CPI seed keywords. One of the most important processing steps is how to construct appropriate prompt information so that the fully pre-trained PANGU model can quickly understand the natural language processing task to be processed. Aiming at the keyword expansion problem of CPI prediction in this paper, the main task of this paper is to extract representative keywords from the obtained CPI-related Baidu Baike texts. Therefore, this article constructs the prompt information as follows:

$$\text{Input Text} = \text{"Keyword Extraction：} \backslash n\text{"} + \text{"Abstract："} + \text{Baike} + \text{"}\backslash n\text{"} + \text{"Keywords："}$$

Here "\n" represents a newline, the purpose is to let the model understand the structure level of the input text. "Keyword Extraction" indicates the type of NLP task that needs to be handled by PANGU model. "Abstract:" and the following "Baike" represent the relevant text of the CPI seed keywords captured from Baidu Baike, which is the text content that the PANGU model needs to focus on. "Keywords:" prompts the PANGU model to output the recognized text keywords, and then extract representative keywords from a large number of Baike texts.

### 4.3. PANGU Robust Test

Due to the randomly initialized neural network parameters and data shuffle, many deep learning models would have unstable performance even under the same hyperparameter setting. Therefore, we test the robustness of our designed keywords extraction method through multiple experiments. The intuition is that we want to make sure that every generated keyword is supposed to be in the result of our multiple experiments. let $A_i$ as the generated keywords in the $i_{th}$ experiments. We define the final generated keywords set $A$ as follows:

$$A = \cap_{i=1}^{N} A_i \tag{12}$$

In our specific CPI keywords generations task, we repeat the keywords generation experiments ten times and found that the PANGU model has a good stable performance with the same keywords output each time. Therefore we prove the stability of PANGU model through our multiple experiments.

# 5. Results

## 5.1. Data Description

Based on the previous literature research, this paper will explore the web search keywords related to CPI from the perspectives of micro and macro indicators. Among them, the micro indicators include eight categories of CPI, and the macro indicators include common economic and financial indicators. The keywords collected from the above perspectives will be used as seed keywords, and then expanded from the perspectives of similarity and importance. The specific seed words are shown in Table 1.

**Table 1.** Macro and micro seed keyword

| Perspectives | Seed Keywords |
|---|---|
| Macro | CPI, GDP, price, product price, price increase, price decrease, deposit rate, economy, currency, capital, inflation, deflation, finance, market, commodity, tax, trade, population |
| Micro | Rice, wheat, pork, beef, fish, vegetables, fruits, edible oil, steamed bread, bread, beverages, liquor, tobacco, men's clothing, women's clothing, children's clothing, rent, transportation, education, medical care, health care, automobiles, gasoline |

According to the above seed keywords, this paper crawls the relevant keyword encyclopedia texts from Baidu Baike, and filters out long paragraph texts with a length of about 300 to 500 words as a corpus for expanding keywords. As a corpus text of popular science and encyclopedia nature, Baidu Baike has high readability and accuracy, so it is suitable for extracting high-quality predictive keywords related to CPI. On this basis, this paper combines the BERT and PANGU models, and performs parallel computing from the perspectives of similarity and importance to determine the similar keywords and the important keywords. Further, we take the intersection of the two to obtain representative keywords that are not only similar to the seed keywords, but also have the ability to summarize and generalize in the context, so as to improve the model's ability to interpret and predict the CPI index.

## 5.2. BERT Keyword Similarity Selection

In the process of selecting CPI-related texts based on similarity, this paper uses the BERT vectorization method to encode and represent the words from JIEBA word segmentation, so as to further calculate the cosine similarity between the seed word and the generated word. According to the cosine similarity between words and a preset selecting threshold, we obtain similar keywords related to seed keywords. Specifically, a total of 166 macro-similar keywords and a total of 400 micro-similar keywords with the similarity of seed words above the threshold of 0.9 were selected in this paper.

## 5.3. PANGU Keyword Importance Selection

This paper introduces the PANGU model from the perspective of importance to extract the keywords from CPI-related Baike text. As a generative model, the PANGU model can integrate the relevant text of CPI and appropriate prompt information, and auto-regressively generate a keyword sequence that could represent the text. At the same time, the PANGU model performs well in zero-shot learning problems, so it can avoid downstream fine-tuning on a specific keyword dataset, thereby preventing keyword extraction bias caused by dataset selection. Finally, a total of 782 macro-important keywords and 1,138 micro-important keywords were obtained.

Considering the similar keywords obtained by BERT similarity screening and the important keywords obtained by PANGU importance screening, this paper takes the intersection of the two as the generated keywords, with potential CPI prediction ability, and obtains its corresponding Baidu search index. Combined with the availability of Baidu search index and the need for CPI forecasting, this paper sets the time interval of the experiment from January 2011 to January 2022. The CPI index is the month-on-month data in the RESSET macro database. The Baidu indexes of the keywords in the generated keywords are the daily data from January 1, 2011, to January 31, 2022, which is converted into monthly data after averaging, and the change rate of the monthly search index is calculated. After these processings, a total of 18 macro seed words, 27 macro generated words, 23 micro seed words, and 76 micro generated words are obtained, as shown in Table 2.

**Table 2.** Macro and micro keyword expansion results

| Perspective | Seed Keywords + Generated Keywords |
|---|---|
| Macro | CPI, GDP, price, product price, price increase, price decrease, deposit rate, economy, currency, capital, inflation, deflation, finance, market, commodity, tax, trade, population, **tax system, business, tax system, population density, Sales price, tax revenue, value-added tax, total investment, PPI, IMF, border trade, paper money, capital, business, tax rate, tax policy, electronic money, finance, customers, money, financial institutions, demand deposits, monetary system, World markets, commodity prices, financial markets, time deposits**[1] |
| Micro | Rice, wheat, pork, beef, fish, vegetables, fruits, edible oil, steamed bread, bread, beverages, liquor, tobacco, men's clothing, women's clothing, children's clothing, rent, transportation, education, medical care, health care, automobiles, gasoline, **higher education, Beef in sauce, shirts, fashionable women's clothing, fish gallbladder, flue-cured tobacco, famous wine, passenger cars, peanut oil, bakery, diesel cars, swine fever, fragrant rice, Xifeng wine, clothing, health, wine, corn oil, tobacco industry, dough, burning Fish, animal oil, common wheat, food, tobacco seeds, green** |

|  | **vegetables, children, whole wheat, green vegetables, porridge, eggs, aged rice, brown bread, fish, model car, marinade, tobacco seedlings, peppers, rice, pig prices, Lamb, shochu, trousers, gasoline for cars, rental, marinated chicken, drinks, sesame oil, pigs, teaching materials, steamed wine, fashion, health, soft drinks, hospitals, nicotine, frozen fish, sesame oil, glutinous rice, fish oil, floating wheat, Clothing, house rental, soybean oil, Ministry of Education, suits, learning, tobacco industry, vegetable oil, automobile industry, dumplings, car brands, renting a house, fruits and vegetables, pig house, Dong wine[2]** |
|---|---|

[1][2] The words marked in bold are generated keywords. Some words are synonyms in Chinese.

### 5.4. Regression Analysis

This paper compares the explanatory power of the original seed keywords and the generated keywords for the CPI prediction through regression analysis. In this paper, the data collected above are processed as follows. First, the CPI consumer price index $CPI_{MoM}$ adopts the month-on-month data (last month = 100), after subtracting the value of the previous month, it represents the change ratio of the price level of household consumer goods in the current month compared with the previous month. Secondly, the Baidu search index of the seed keywords and the expanded keywords represents the frequency of words appearing in online searches in that month, and reflects the residents' attention to word-related events. Therefore, this paper also uses the month-on-month data of the Baidu search index, and calculates the change ratio of the search index after subtracting the search index of the previous month, which is more conducive to analyzing the impact of changes in the attention of CPI-related words on CPI changes. Finally, considering the possible influence of an annual cycle, this paper performs lag processing of 1 to 12 orders on the CPI change rate and the Baidu search index change rate, to tap the potential lag forecasting ability.

At the same time, because the Baidu search index represents the degree of residents' attention to specific concepts, it will be affected by other factors other than price changes, and there will be a lot of noise. This paper uses the Pearson correlation coefficient to remove the keyword variables that are too low in the CPI change rate, and then removes the impact of other factors other than price changes on the Baidu search index. The absolute threshold of the correlation coefficient is 0.3, and the variables with the correlation coefficient with the CPI change rate below 0.3 are eliminated. In this paper, the above operations are performed on the seed words and the generated words respectively, and finally a total of 71 predictors under different lag orders are obtained.

Due to the large number of seed words and generated words, and multi-order lag processing at the same time, the above predictors contain many time series. Therefore, in this paper, principal component analysis method is used to reduce the dimension of the predictor variables, and finally six principal components are obtained, which can explain at least 70% of the original variables. Finally, six principal components were determined as independent variables, and the rate of change of CPI was used as the dependent variable, and the principal component regression results were shown in Table 3.

**Table 3.** CPI regression results

|  | Seed Keywords | | | Generate Keywords | | |
|---|---|---|---|---|---|---|
| Principal Components | Component Number | Regression coefficients | Cumulative Proportion | Component number | Regression coefficients | Cumulative Proportion |
|  | 1 | 0.0077*** | 0.3098 | 1 | 0.0032*** | 0.3574 |
|  | 2 | 0.0024** | 0.5061 | 2 | -0.0040*** | 0.5010 |
|  | 3 | 0.0061*** | 0.6214 | 3 | 0.0017** | 0.5947 |
|  | 4 | 0.0024 | 0.7179 | 4 | 0.0037*** | 0.6623 |
|  | 5 | 0.0007 | 0.7677 | 5 | -0.0022* | 0.6991 |
|  | 6 | 0.0034 | 0.8096 | 6 | -0.0003 | 0.7348 |
| F statistic | 17.63 | | | 21.23 | | |
| DW statistics | 1.470 | | | 1.868 | | |
| $R^2$ | 0.481 | | | 0.528 | | |
| sample size | 120 | | | 120 | | |

Note: *, **, *** represent the 10%, 5%, and 1% significance levels, respectively

Significance levels of the model F statistics after the principal component analysis are all below 1%, indicating that the regression fitting effects of the model on CPI are good. At the same time, this paper uses R-square to compare the predictive and explanatory power of the model. The results show that after adding the generated keywords, the corresponding R-squared is still higher even though the cumulative PCA explanation proportion of the first six principal components is lower. Specifically, the R-square of the model corresponding to the original keyword is 0.481, while the R-square of the model corresponding to the expanded keyword after filtering by BERT similarity and PANGU importance is 0.528. On the other hand, the significance of the regression coefficients of each principal component of the generated keywords has been improved to varying degrees. The above empirical analysis shows that the generated keywords obtained by natural language processing technology based on similarity and importance in this paper can improve the prediction effect of the model on the CPI change rate.

## 6. Conclusions

This paper introduces the BERT model and PANGU model in natural language processing to expand the predictor variables in the CPI prediction problem. Among them, the Bert vectorization model is based on the Transformer encoder structure, which encodes words into word vectors, and mines keywords similar to the seed keywords from the perspective of similarity. The PANGU model is based on the Transformer decoder structure, and considers the global semantics of the sentence to autoregressively generate keywords that have the ability to represent context. Through the parallel calculation of the two, this paper extracts the final expanded keywords, and uses the principal component regression to compare the original seed keywords and the final generated keywords respectively, and verifies the effect of the generated keywords' CPI prediction ability.

First of all, the Bert vectorization model used in this paper can filter and match the CPI-related Baidu Baike text after JIEBA word segmentation from the perspective of similarity by encoding a single word into a vector form. The multi-head self-attention mechanism structure and feed-forward neural network structure in the encoder based on the Transformer structure can effectively consider the deep semantics of words, so as to get better vector representation of keywords and obtain similar keywords related to CPI.

Secondly, as a super-large Chinese pre-trained language model, the PANGU model can get rid of the dependence on the downstream fine-tuning data set under the appropriate prompt information, and automatically generate keyword sequences by integrating the context, effectively avoiding the selection of fine-tuning data sets. Therefore, we can generate the important keywords with a stronger representation ability of the context.

Finally, the parallel keyword expansion method of similarity and importance adopted in this paper can mine keywords with better predictive ability for CPI index. Through principal component regression analysis, this paper conducts an empirical test on the prediction effect of the CPI change rate of the seed keywords and the generated keywords. The empirical results show that the finally generated keywords have better predictive ability for CPI than the seed keywords. Therefore, the text mining process based on the BERT vectorization model and the PANGU model designed in this paper can effectively extract high-quality keywords with both economic meaning and stronger predictive ability from the Baike text for CPI prediction analysis.

# References

[1] Zhang Y. Short-term CPI forecast in the context of big data[J]. *China Statistics*, **2015**(12):49-50. (In Chinese)

[2] Yuan M. A Nowcast Model of CPI Based on Search Volume of Online Shopping[J]. *Journal of Statistics and Information*, **2015**, 30(04):20-27. (In Chinese)

[3] Dong L, Peng K Y, Tang X B. Research on real-time CPI prediction in the context of big data[J]. *The World of Survey and Research*, **2017**(08):51-54. (In Chinese)

[4] Xu Y M, Gao Y M. Construction of the Public Opinion Index of CPI based on the Internet Big Data [J]. *The Journal of Quantitative and Technical Economics*, **2017**,34(01):94-112. (In Chinese)

[5] Zhang H, Shen H L, Xia L. Research on CPI Prediction based on Multi-source Asynchronous Mixed Sampling Data[J]. *The Journal of Quantitative and Technical Economics*, **2020**, 37(10):149-168. (In Chinese)

[6] Vaswani A, Shazeer N, Parmar N, et al. Attention is all you need[J]. *Advances in neural information processing systems*, **2017**, 30.

[7] Tang X, Dong M R, Xu R. Design and Application of Text Mining Technology for CPI Prediction Based on Big Data[J]. Statistical Research, 2021, 38(08):146-160. (In Chinese)

[8] Zeng W, Ren X, Su T, et al. PanGu-$\alpha$: Large-scale Autoregressive Pretrained Chinese Language Models with Auto-parallel Computation[J]. arXiv preprint arXiv:2104.12369, 2021.

[9] Mindspore, 2020, https://www.mindspore.cn/.

[10] Brown T, Mann B, Ryder N, et al. Language models are few-shot learners[J]. *Advances in neural information processing systems*, **2020**, 33: 1877-1901.

[11] Devlin J, Chang M W, Lee K, et al. Bert: Pre-training of deep bidirectional transformers for language understanding[J]. *arXiv preprint arXiv:1810.04805*, **2018**.

[12] Radford A, Narasimhan K, Salimans T, et al. Improving language understanding by generative pre-training[J]. **2018**.